\documentclass{ws-ijmpc}
\usepackage[dvips]{color}

\begin{document}

\markboth{Authors' Names}
{Complex network analysis of literary and scientific texts}

\catchline{}{}{}{}{}
%%%%%%%%%%%%%%%%%%%%%%%%%%%%%%%%%%%%%%%%%%%%%%%%%%%%%%%%%%%%%%%%%%%%

\title{COMPLEX NETWORK ANALYSIS OF LITERARY AND SCIENTIFIC TEXTS}

\author{IWONA GRABSKA-GRADZI\'NSKA}

\address{Faculty of Physics, Astronomy and Applied Computer Science, Jagiellonian University,\\ ul.~Reymonta~4, 30-059 Krak\'ow, Poland\\
grabska@gmail.com}

\author{ANDRZEJ KULIG}

\address{Institute of Nuclear Physics, Polish Academy of Sciences,\\
ul. Radzikowskiego 152, 31-342 Krak\'ow, Poland\\
andrzej.kulig@ifj.edu.pl}

\author{JAROS{\L}AW KWAPIE\'N}

\address{Institute of Nuclear Physics, Polish Academy of Sciences,\\
ul. Radzikowskiego 152, 31-342 Krak\'ow, Poland\\
jaroslaw.kwapien@ifj.edu.pl}

\author{STANIS{\L}AW DRO\.ZD\.Z}

\address{Institute of Nuclear Physics, Polish Academy of Sciences,\\
ul. Radzikowskiego 152, 31-342 Krak\'ow, Poland\\
Faculty of Physics, Mathematics and Computer Science, Cracow University of Technology,\\
ul. Warszawska 24, 31-155 Krak\'ow, Poland\\
stanislaw.drozdz@ifj.edu.pl}

\maketitle

\begin{history}
\received{Day Month Year}
\revised{Day Month Year}
\end{history}

\begin{abstract}

We present results from our quantitative study of statistical and network properties of literary and scientific texts written in two languages: English and Polish. We show that Polish texts are described by the Zipf law with the scaling exponent smaller than the one for the English language. We also show that the scientific texts are typically characterized by the rank-frequency plots with relatively short range of power-law behavior as compared to the literary texts. We then transform the texts into their word-adjacency network representations and find another difference between the languages. For the majority of the literary texts in both languages, the corresponding networks revealed the scale-free structure, while this was not always the case for the scientific texts. However, all the network representations of texts were hierarchical. We do not observe any qualitative and quantitative difference between the languages. However, if we look at other network statistics like the clustering coefficient and the average shortest path length, the English texts occur to possess more clustered structure than do the Polish ones. This result was attributed to differences in grammar of both languages, which was also indicated in the Zipf plots. All the texts, however, show network structure that differs from any of the Watts-Strogatz, the Bar\'abasi-Albert, and the Erd\"os-R\'enyi architectures.

\end{abstract}

\ccode{PACS Nos.: 64.60aq, 89.75Fb}

\section{Introduction}

Natural language is an evolving system whose present structure can doubtlessly be considered a product of long history of self-organization~\cite{christiansen08}. Like for many other self-organized systems known in Nature, the observables associated with language, being, for example, written texts or spoken messages, reveal quite sophisticated dynamics. Any language sample by no means is an amorphous mixture of symbols (letters, phonemes, morphemes, words, etc.) but rather a highly organized sequence in which particular symbols are ordered according to specific rules most of which are defined by the language grammar. Since the existence of grammar is an emergent phenomenon~\cite{nowak99,nowak00}, language can be counted among the complex systems~\cite{anderson72,kwapien12}. The grammatical rules together with the information content impose on the language elements relations which can be most easily expressed in a form of network where, for instance, words are expressed by nodes and their relations by edges. Some earlier attempts along this way were presented in Refs.~\cite{masucci06,caldeira06,costa07,zhou08,roxas12} for English, Portuguese and Chinese. Here we show a few results that were obtained for English and Polish and for different types of texts (literary or scientific).

\section{Methods and data}

Our analysis was based on texts samples written in two languages: English and Polish. Both belong to the Indo-European family, but to different groups: the West-Germanic and the West-Slavic group, respectively. Their grammar therefore significantly differs, most notably in the existence of rich inflection of words in Polish as compared to a rather residual one in English. However, in the present work we do not deal with the semantic analysis, but restrict our study to a statistical analysis of word adjacency. As regards the English part, we analyze two groups of texts. The first one comprises the literary texts represented by 9 works of prose (``Ulysses'' and ``Finnegans Wake'' by J.~Joyce, ``Alice's Adventures in Wonderland'' by L.~Carroll, ``Adventures of Huckelberry Finn'' by M.~Twain, ``Pride and Prejudice'' by J.~Austen, ``Oliver Twist'' by C.~Dickens, ``Secret Adversary'' by A.~Christie, ``Adventures of Sherlock Holmes'' and ``Study in Scarlet'' by A.~Conan Doyle), 4 dramas by W.~Shakespeare (``Hamlet'', ``Macbeth'', ``Winter Tale'', and ``Romeo and Juliet''), 61 poems by O.~Wilde, and 25 poems by T.S.~Elliott. The second group comprises the scientific texts represented by selected works of E.~Witten~\cite{witten.list}, E.G.D.~Cohen~\cite{cohen.list}, S.~Weinberg~\cite{weinberg.list}, and P.W.~Anderson~\cite{anderson.list}, as well as by three long reviews by D.~Sornette~\cite{sornette03}, R.~Albert and A.-L.~Barab\'asi~\cite{albert02}, and J.~Kwapie\'n and S.~Dro\.zd\.z~\cite{kwapien12}. Somewhere at the interface of these two groups, there is ``A Brief History of Time'' by S.~Hawking and ``The Emperor's New Mind: Concerning Computers, Minds, and the Laws of Physics'' by R.~Penrose representing popular science. The Polish language was represented by the novels: ``Lalka'' (``The Doll'') by B.~Prus, ``Bramy Raju'' (``The Gates of Paradise'') by J.~Andrzejewski, ``Dolina Issy'' (``The Issa Valley'') by C.~Mi{\l}osz, ``Cesarz'' (``The Emperor: Downfall of an Autocrat'') by R.~Kapu\'sci\'nski, ``Dzienniki gwiazdowe'' (``The Star Diaries'') by S.~Lem, ``Ferdydurke'' by W.~Gombrowicz, the epic ``Pan Tadeusz'' (``Sir Thaddeus'') by A.~Mickiewicz, the only Polish translation of ``Ulysses'' done by M.~S{\l}omczy\'nski, 35 poems by C.~Mi{\l}osz, and 99 poems by W.~Szymborska.

All the texts were filtered in order to remove some of the punctuation marks (all except the ones that can functionally end sentences: the periods, the colons and semicolons, the question and exclamation marks) as well as the non-word sequences like numbers. The so-preprocessed texts were subject to further analysis.

\section{Results}

Although language and language samples are traditionally subject to purely qualitative analysis in the fields of humanities, the language samples consist of symbolic sequences, which can be easily subject to quantitative analysis. Historically, the beginning of quantitative analysis of natural language is usually associated with the name of G.K.~Zipf, who was the first to carry out an extensive study of word frequencies in written texts in a few different languages~\cite{zipf32,zipf49}, despite that in fact he also had known predecessors like J.-B.~Estoup~\cite{estoup16} and E.L.~Thorndike~\cite{thorndike32} who did some research in the same direction but far less extensive than the Zipf's and without any significant impact on science. The main result attributed to Zipf is his eponymous law stating that number $F$ of occurrences of words ordered according to their relative frequency in text samples is, roughly, inverse proportional to their rank $R$. For the English language, it is:
\begin{equation}
F(R) = {A \over R^{\alpha}}, \qquad \alpha \approx 1.
\label{eqn::zipf}
\end{equation}
while for other languages $\alpha$ can also be slightly smaller or larger than 1. $A$ stands here for an empirical proportionality constant equal to $0.1 T$, where $T$ is the total number of words in a sample (a sample's length). For single texts samples like books, the above power-law relation is usually well preserved for medium ranks (e.g., $10 < R < 1000$) in the majority of cases, but for the lowest and the highest ranks it breaks down leading to flattening of $F(R)$ for the most frequent words and to its faster decline for the least frequent ones. As regards the multi-piece unions of text samples (corpora), the scaling~(\ref{eqn::zipf}) holds up to the ranks of a few thousand and above them another scaling region with a larger value of $\alpha$ appears~\cite{montemurro01,ferrer01}. The latter phenomenon can be interpreted as a division of the vocabulary into two sections: the basic vocabulary containing words that are shared by almost all the text samples, thus common to all people, and the specialist vocabulary that can be subject-specific and author-specific~\cite{montemurro01}. An interesting property of the Zipf law is its similarity to a number of other relations that can be found in different and sometimes very distant fields: city population, scientific paper citations, earthquake magnitude, and many more~\cite{newman05}.

\begin{figure}
\centerline{\psfig{file=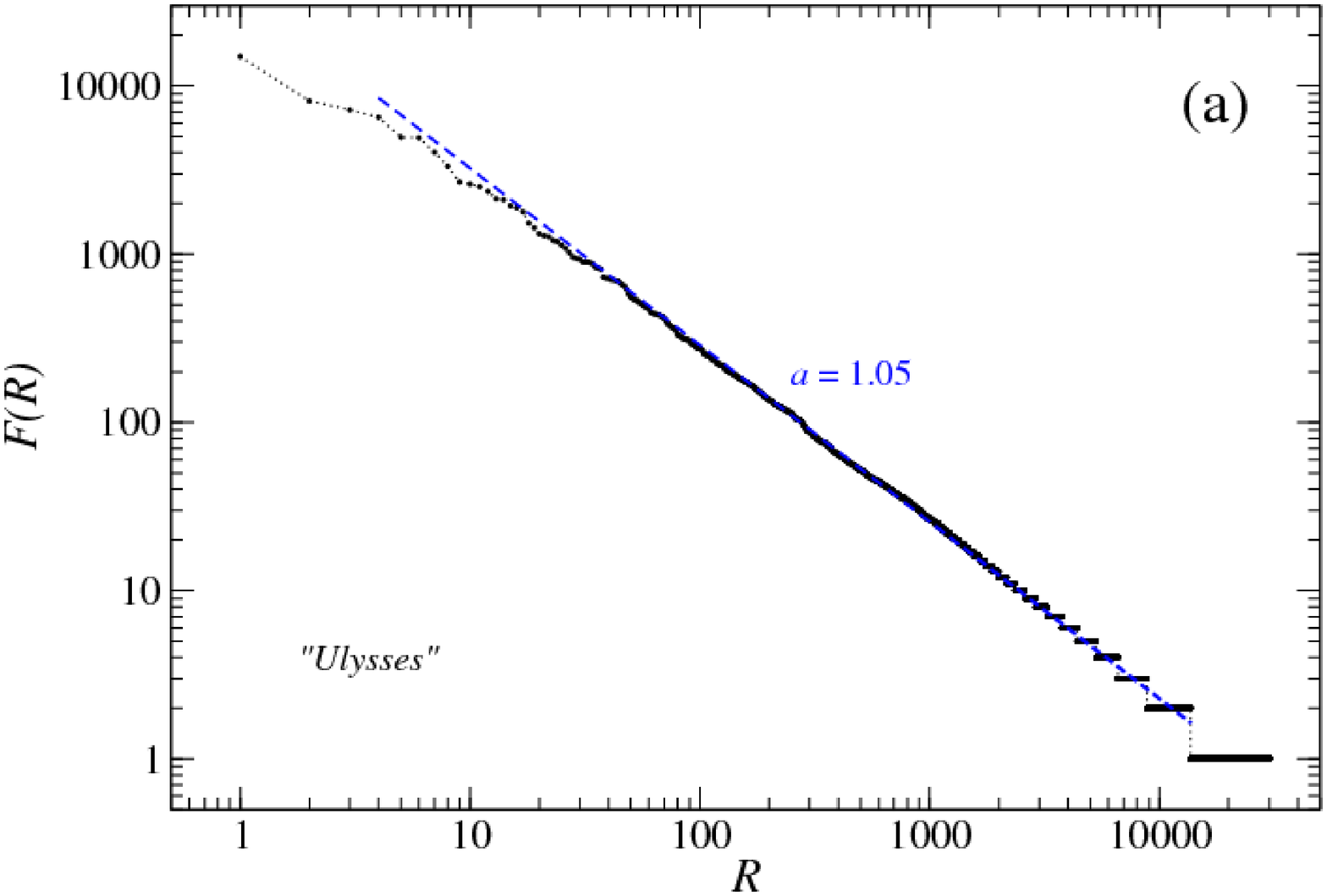,width=5cm}\psfig{file=fig1b.eps,width=5cm}}

\vspace{0.2cm}
\centerline{\psfig{file=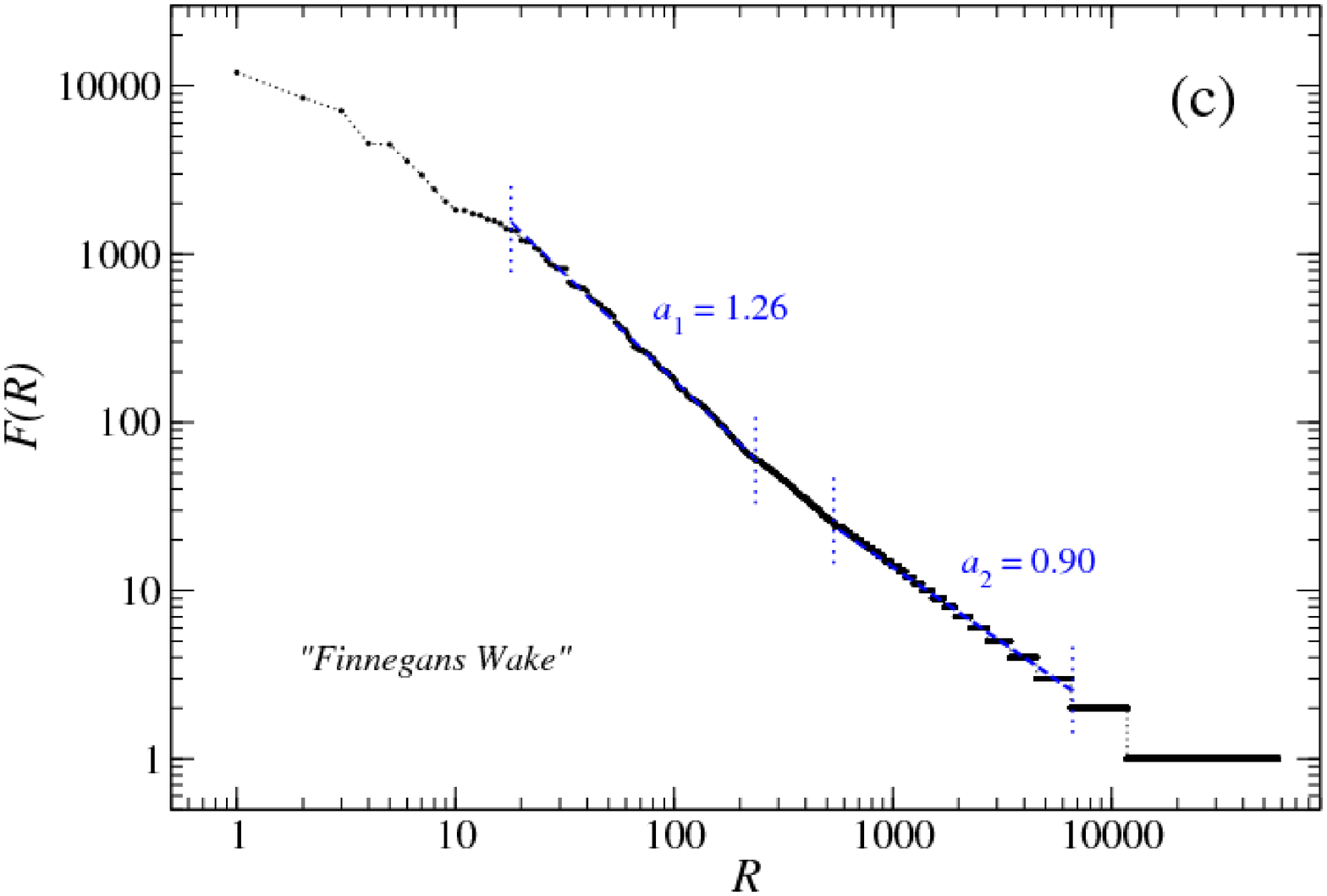,width=5cm}\psfig{file=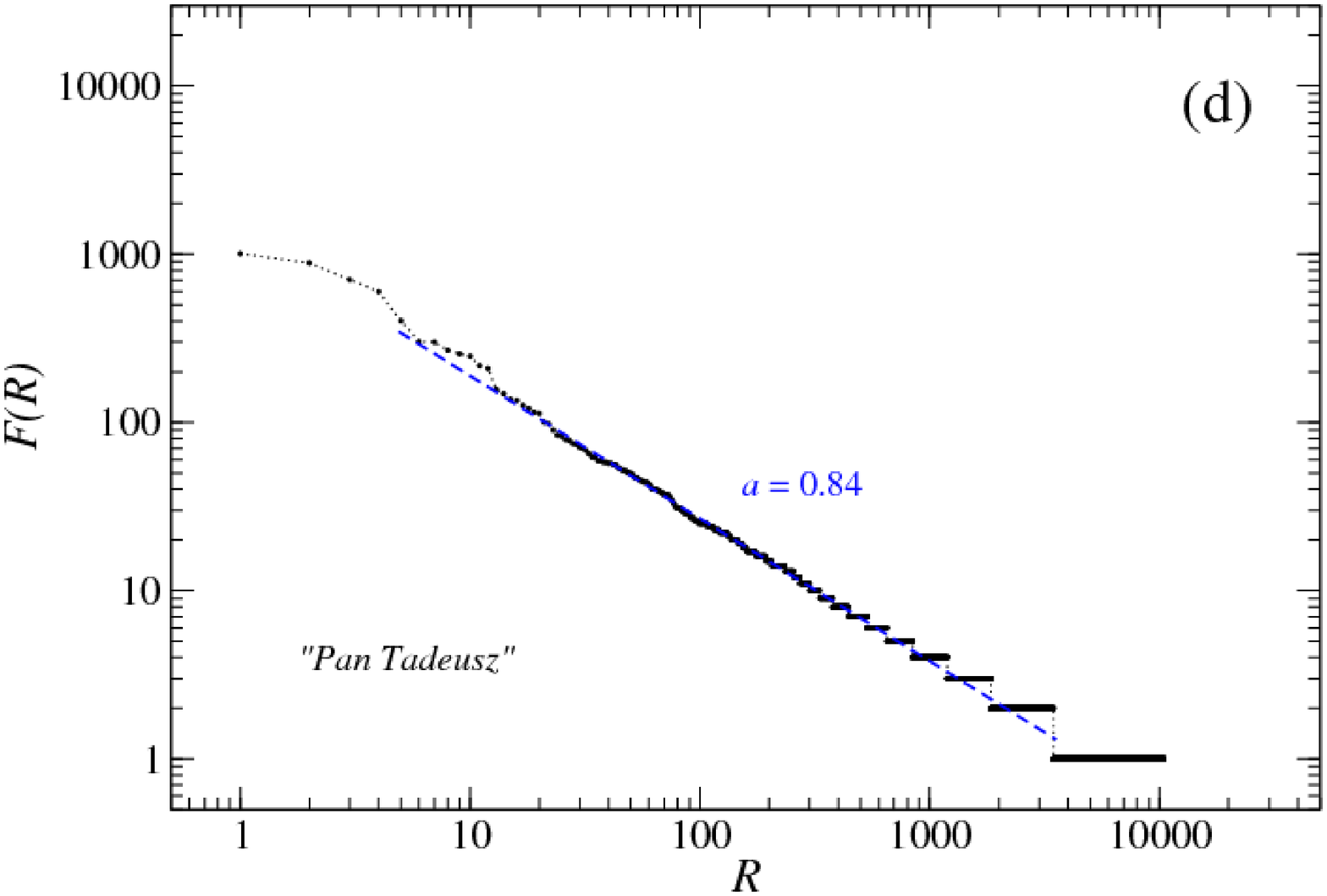,width=5cm}}
\caption{Rank-frequency distribution of words $F(R)$ in the English texts of (a) ``Ulysses'' by J.~Joyce, (b) ``A Brief History of Time'' by S.~Hawking, (c) ``Finnegans Wake'' by J.~Joyce, and (d) in the Polish text of ``Pan Tadeusz'' (``Sir Thaddeus'') by A.~Mickiewicz. The empirical distributions are compared with the corresponding best fits in terms of a power-law function (Eq.~(\ref{eqn::zipf})) with the scaling exponent $\alpha$.}
\label{fig::zipf}
\end{figure}

As regards the rank-frequency relation for text samples, Fig.~\ref{fig::zipf}(a) shows such a plot for ``Ulysses''. It is notable for its uniquely broad range of ranks (3 decades) for which a power-law scaling holds, which is extremely rarely equalled by other pieces of texts. For example, a similar plot for ``A Brief History of Time" in Fig.~\ref{fig::zipf}(b) reveals scaling valid for only 2 decades. This means that the vocabulary volume of this book is smaller than it would be expected from the power-law relation holding over all the ranks. On contrary, ``Finnegans Wake'' (Fig.~\ref{fig::zipf}(c)) possesses extremely diverse vocabulary and the rarest words are overrepresented leading to breaking of the Zipf-like relation in the opposite direction as compared to the Hawking's book. Such a situation does not surprise us, however, since ``Finnegans Wake'' is known to be a highly experimental piece of text comprising words from many languages. Next, Fig.~\ref{fig::zipf}(d) shows a rank-frequency plot obtained for the Polish text of ``Pan Tadeusz''. A well-fitted power-law function with $\alpha = 0.84$ describes the plot, whose slope is much smaller than for typical English texts and even for typical Polish ones ($\alpha = 0.94$~\cite{kwapien09,kwapien10}), while it is characteristic for the works of A.~Mickiewicz~\cite{orczyk08}.

Characterization of a text sample by means of the Zipf plot is informative in respect to vocabulary volume of an author and mutual relations between the most common and other words, but it is insensitive to any kind of correlations possibly present in the sample. In order to incorporate correlations into our analysis, we create network representations of each of the text sam2ples studied here. We choose such a representation in which different words are regarded as different network nodes. One type of interesting correlations that can be quantified in this way is the adjacency relation between pairs of words. Two words are considered related and their nodes linked by an edge if they are the nearest neighbours at least once in a sample. For a given word, there are two possible relations with its neighbours: precedence and succession. The former is when a neighbour precedes the considered word, while the latter is in the opposite case. In this context, we study two networks: the precedence (left-side neighbourhood) network and the succession (right-side neighbourhood) network. We decided to consider only the neighbours that belong to the same sentence and neglect the inter-sentence adjacency. This does not influence our results, however: a preliminary analysis carried on a few text samples showed that there is no qualitative difference of the results between these cases. This, of course, might be a consequence of a much smaller number of such inter-sentence pairs (roughly, less than 10\% of all pairs).

\begin{figure}
\centerline{\psfig{file=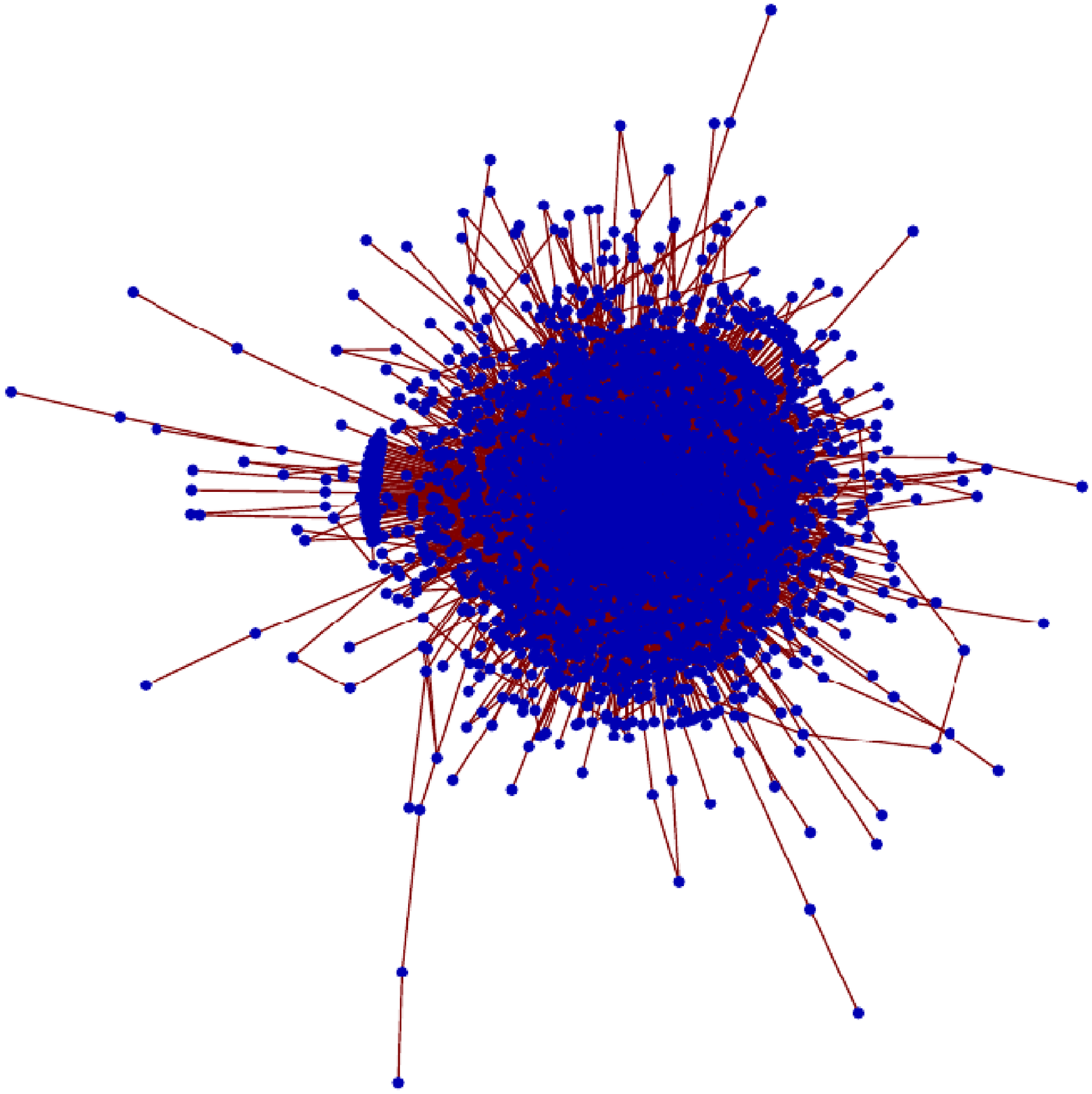,width=5cm}\psfig{file=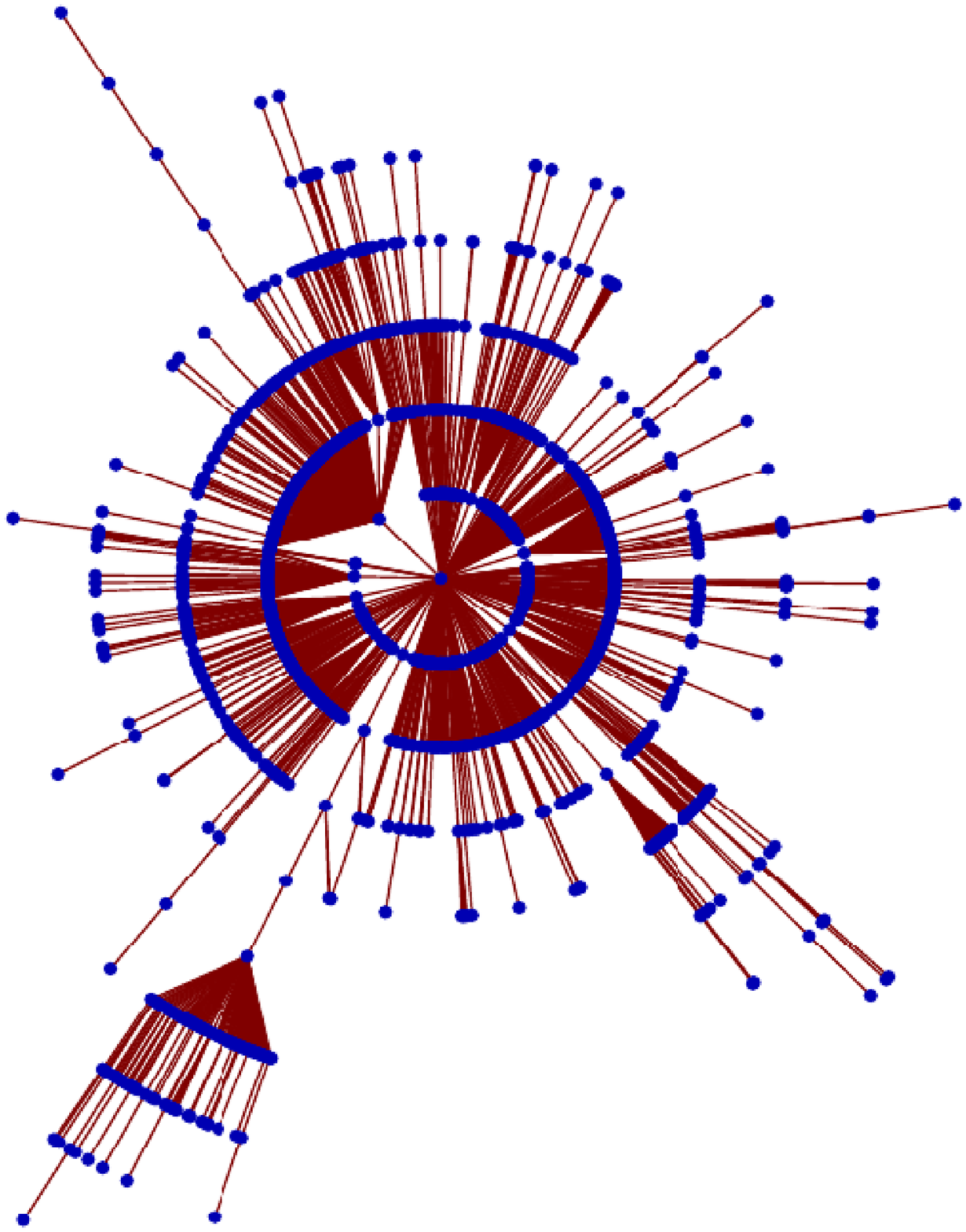,width=5cm}}
\caption{Binary network representation (left) and minimal spanning tree (right) of an exemplary text sample: ``Statistical mechanics of complex networks''. Both pictures correspond to the succession network.}
\label{fig::hedgehog}
\end{figure}

\begin{figure}
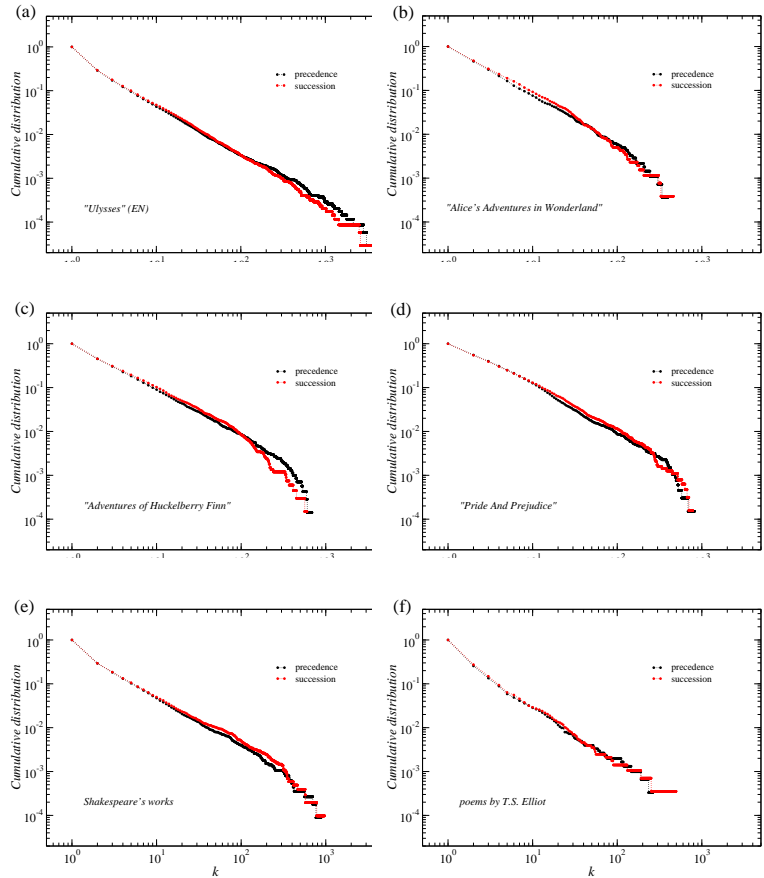

\centerline{\psfig{file=fig3a.eps,width=5cm}\psfig{file=fig3b.eps,width=5cm}}

\vspace{0.2cm}
\centerline{\psfig{file=fig3c.eps,width=5cm}\psfig{file=fig3d.eps,width=5cm}}

\vspace{0.2cm}
\centerline{\psfig{file=fig3e.eps,width=5cm}\psfig{file=fig3f.eps,width=5cm}}
\caption{Cumulative distributions $P(X \ge k)$ of the node degrees $k$ for the word-adjacency network representations of English literary texts: (a) ``Ulysses'', (b) ``Alice's Adventures in Wonderland'', (c) ``Adventures of Huckelberry Finn'', (d) ``Pride and Prejudice'', (e) 4 Shakespeare dramas, and (f) 25 poems by T.S.~Elliott. The precendence and succession networks are shown simultaneously in each panel.}
\label{fig::literary.english.node.degrees}
\end{figure}

By construction, our networks can be either binary or weighted. In the former case, we consider two nodes to be linked by an edge if the respective words are neighbours at least once in a text but we do not pay attention to how many times they neighbour each other. In contrast, in the latter case, we may count the number of such occurrences and attribute a corresponding weight to each edge. Examples of both situations are presented in Fig.~\ref{fig::hedgehog}, where a binary network (left) and a minimal spanning tree calculated from a weighted network (right) are created for an exemplary piece of text.

To begin with, let us calculate a cumulative distribution $P(X \ge k)$ of node degrees $k$ for different texts. This is one of the most informative quantities since it allows one to detect a hierarchical and scale-free structure of a given network~\cite{boccaletti06}. Since such distributions for weighted networks carry basically the same information as the Zipf plots, we restrict our analysis to the binary networks only. Fig.~\ref{fig::literary.english.node.degrees} exhibits cumulative distributions $P(X \ge k)$ for selected literary texts in English. Interestingly, although there are clear differences between the distributions for different texts, all the texts studied (including other not shown here) reveal the scale-free or almost scale-free dependence for some range of $k$, with the scaling exponents $1 < \beta < 2$ being in agreement with the results from other studies of different systems~\cite{redner98,barabasi99}. It happens for some texts that the precedence and the succession networks visibly differ from each other. We do not inspect this issue in more detail, but a source of this difference might be either author-specific or text-specific.

\begin{figure}
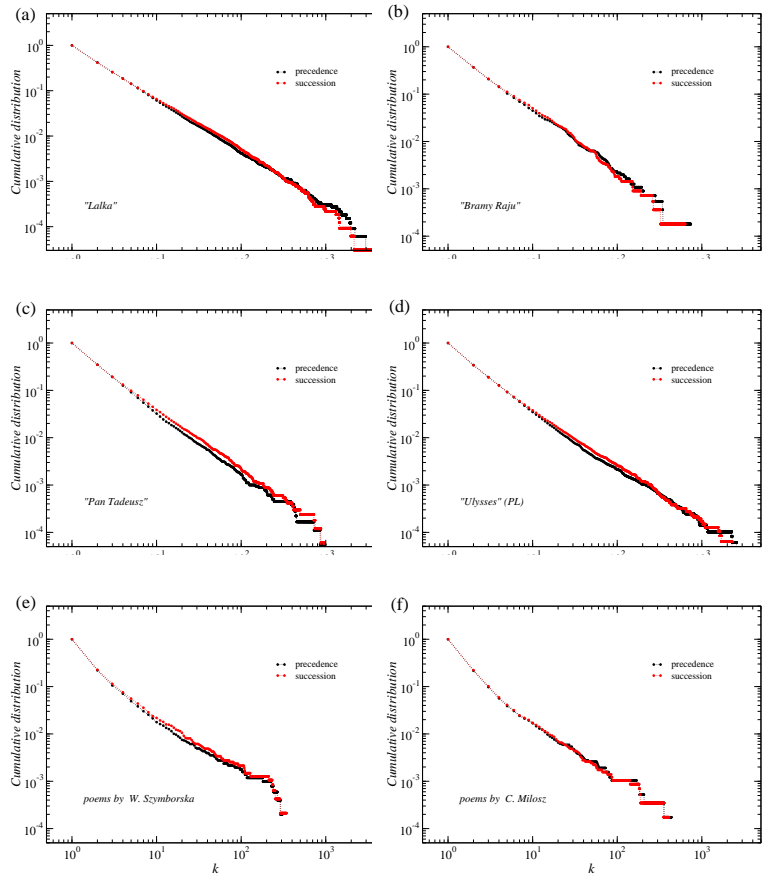

\centerline{\psfig{file=fig4a.eps,width=5cm}\psfig{file=fig4b.eps,width=5cm}}

\vspace{0.2cm}
\centerline{\psfig{file=fig4c.eps,width=5cm}\psfig{file=fig4d.eps,width=5cm}}

\vspace{0.2cm}
\centerline{\psfig{file=fig4e.eps,width=5cm}\psfig{file=fig4f.eps,width=5cm}}
\caption{Cumulative distributions $P(X \ge k)$ of the node degrees $k$ for the word-adjacency network representations of Polish literary texts: (a) ``Lalka'', (b) ``Bramy Raju'', (c) ``Pan Tadeusz'', (d) a Polish translation of ``Ulysses'', (e) 99 poems by W.~Szymborska, and (f) 35 poems by C.~Mi{\l}osz. The precendence and succession networks are shown simultaneously in each panel.}
\label{fig::literary.polish.node.degrees}
\end{figure}

In Fig.~\ref{fig::literary.polish.node.degrees}, the $P(X \ge k)$ distributions are shown for selected Polish literary texts. For prose ((a)-(d)), the scale-free slopes of these distributions are even better visible than for the English texts in Fig.~\ref{fig::literary.english.node.degrees}. The same refers to poems except the overrepresentation of nodes with small $k$ in the case of Polish poetry (Fig.~\ref{fig::literary.polish.node.degrees}(e)-(f)). This overrepresentation probably stems for the fact that poetry, which needs a specific rhythm, imposes strong restrictions on the words that can be used in particular places.

\begin{figure}
\centerline{\psfig{file=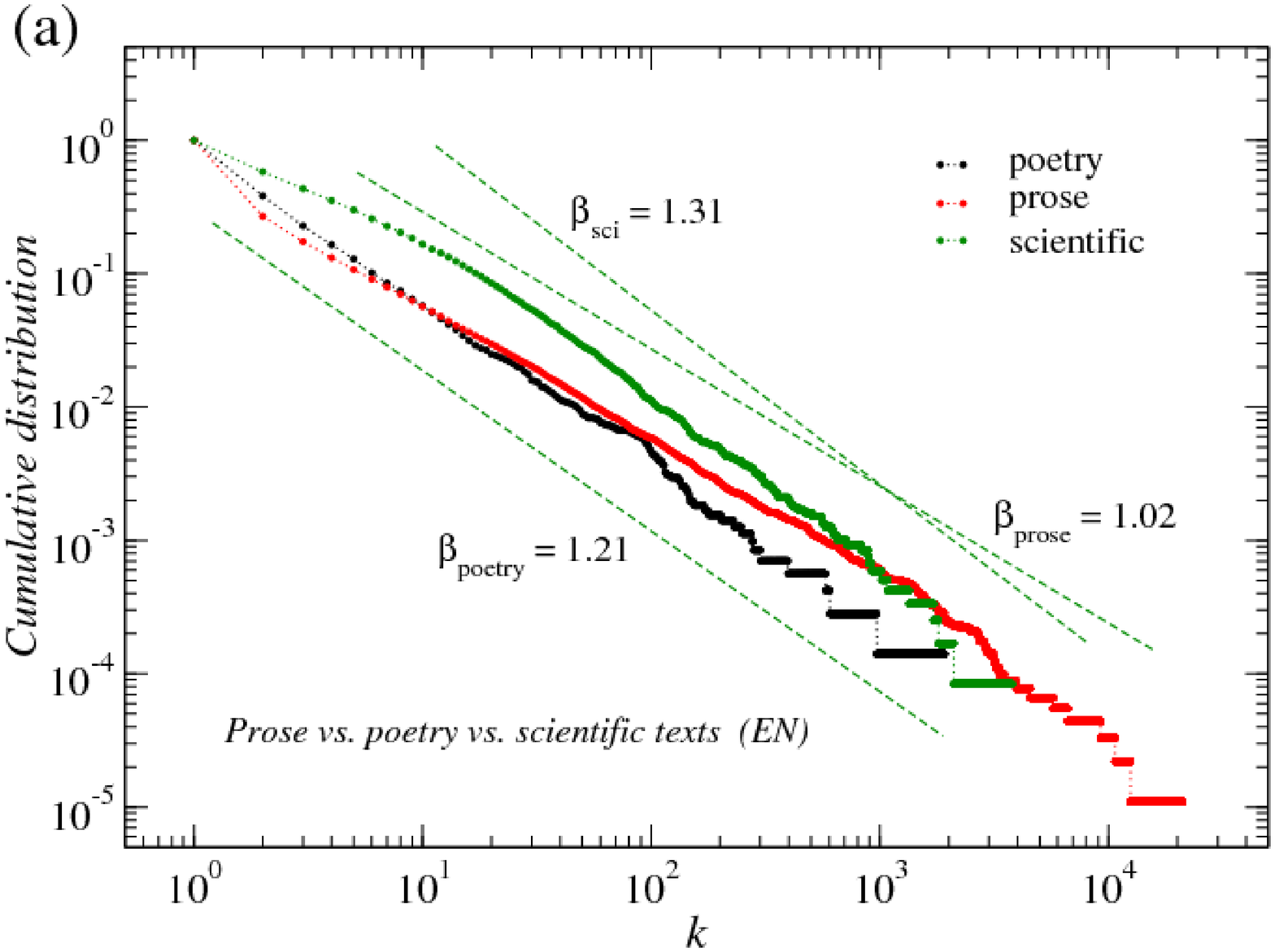,width=10cm}}

\vspace{0.2cm}
\centerline{\psfig{file=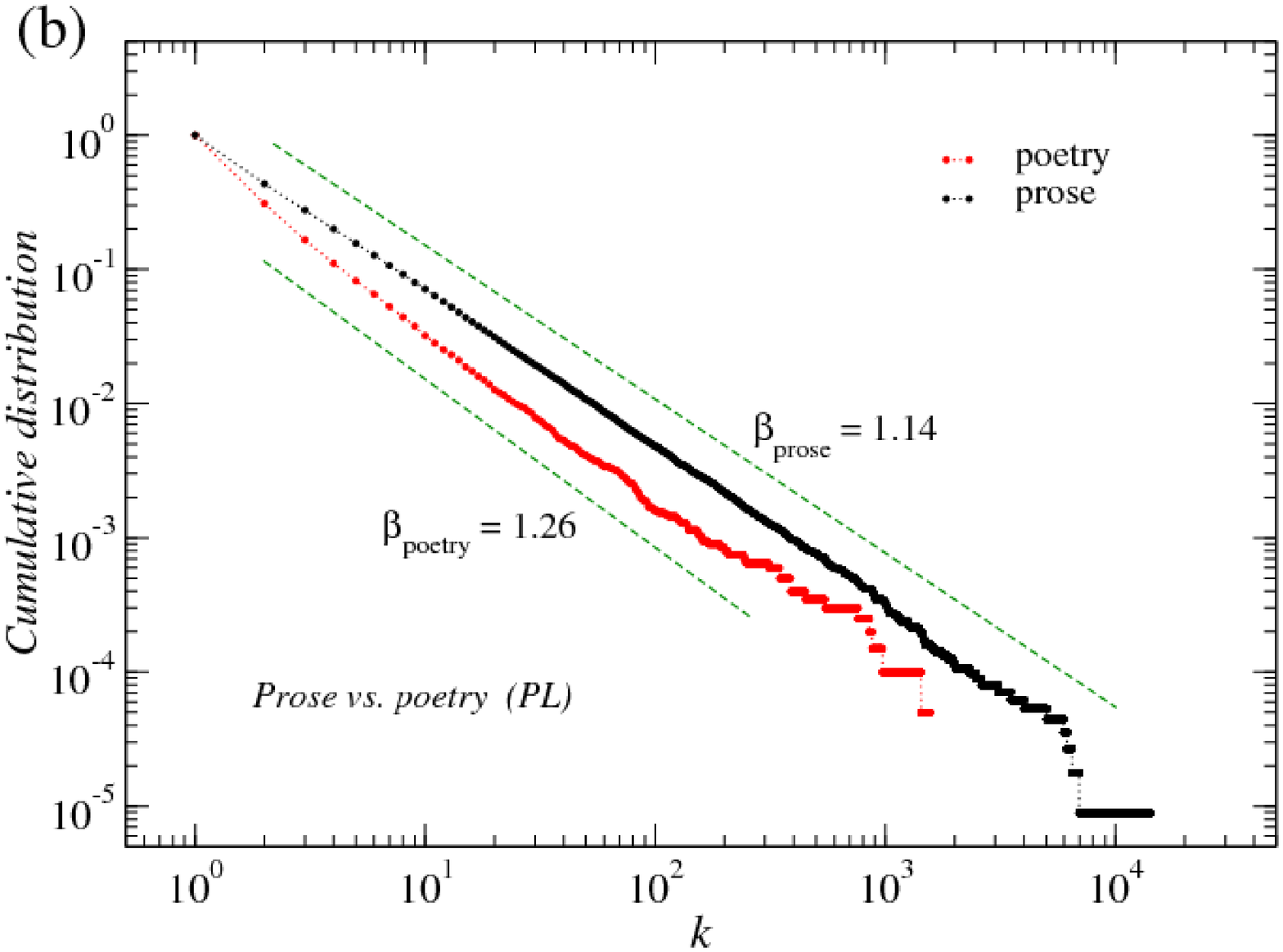,width=10cm}}
\caption{Cumulative distributions $P(X \ge k)$ of the node degrees $k$ for the word-adjacency network representations of English (a) and Polish (b) corpora containing prose, poetry or scientific texts. Slopes of the best-fitted power laws are indicated by the dashed lines and the values of the scaling exponent $\beta$.}
\label{fig::corpora.node.degrees}
\end{figure}

In order to compare statistical properties of node degrees for different types of texts and the two languages, from the texts considered in this work, we create five separate corpora containing English prose, English poetry, English scientific texts, Polish prose, and Polish poetry. Then for each corpus, we calculate a node degree cumulative distribution $P(X \ge k)$. Fig.~\ref{fig::corpora.node.degrees} shows these distributions along with their power-law slopes. Regarding prose, the distribution for the English language has smaller slope than its counterpart for the Polish language. However, the picture for poetry looks different: the node degree distribution is steeper in the case of English. Roughly, as regards the corpora of Polish prose and of Polish poetry, the distributions look similar, which is not the case for their English counterparts. $P(X \ge k)$ for the corpus of scientific texts written in English reveals the most steep slope with $\beta = 1.31$. This is not surprising, however, since many scientific texts are full of mathematics and related formal names and expressions, which make the vocabulary poorer than in the case of literary works, which do not have any vocabulary restrictions (see Fig.~\ref{fig::corpora.node.degrees}(a)). As regards $P(X \ge k)$ for the individual scientific papers, some of them do not reveal any trace of scaling while other are clearly scale-free. This strongly depends on the relative amounts of standard description and strict mathematical language: the less mathematics is there, the better scaling can be observed.

\begin{figure}
\centerline{\psfig{file=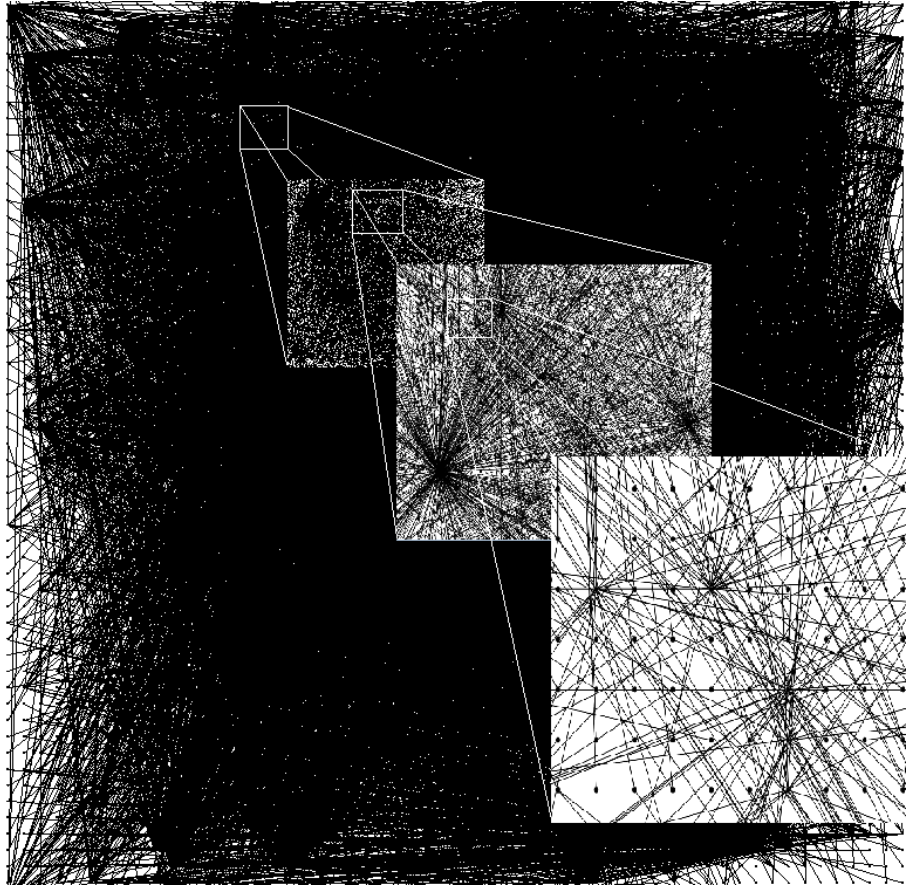,width=12cm}}
\caption{Visualization of a network representation of an exemplary text sample showing a hierarchy of nodes.}
\label{fig::hierarchy}
\end{figure}

\begin{figure}
\centerline{\psfig{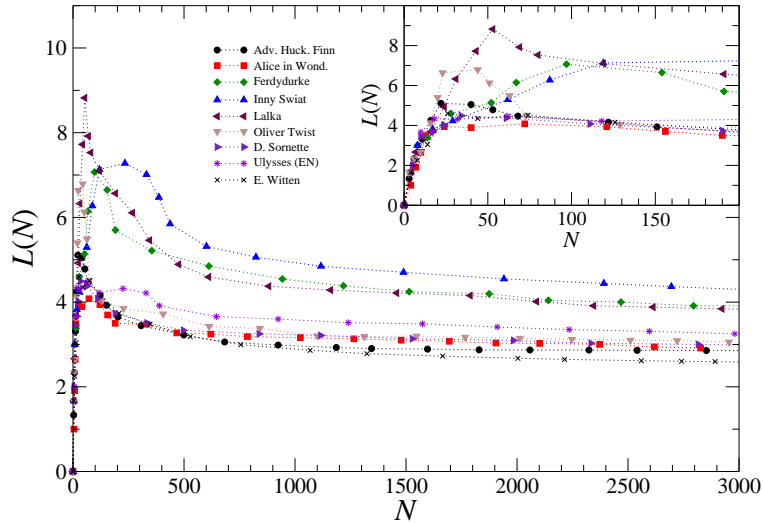}}
\caption{The average shortest path length $L$ as a function of the number of nodes $N$ for exemplary texts considered in this work (only $N < 3000$ is shown because the texts differ in their vocabulary volume). The inset shows magnification of the small-$N$ region.}
\label{fig::path.nodes}
\end{figure}

Our results for both languages indicate that the adjacency networks in both representations are strongly non-democratic with a clear hierarchy of hubs. Indeed, Fig.~\ref{fig::hierarchy} confirms this conclusion by showing both global and local hubs with large values of $k$ surrounded by clouds of peripheral nodes with $k \approx 1$. Other topological properties of the networks can be characterized by their spatial extension and inclination of nodes to form clusters. The former can be quantitatively described by the average characteristic path length $L$ expressing the average node-to-node distance. For a binary network it is defined by:
\begin{equation}
L = {1 \over N (N-1)} \sum_{i,j \in \mathcal{N} \ i \neq j} L_{ij},
\end{equation}
where $L_{i,j}$ is the length of the shortest path connecting the nodes $i$ and $j$ (the length of each path is defined as a number of edges this path passes through). Magnitude of $L$ and, especially, its dependence on $N$ is different for different network types. It typically grows fast for regular lattices and chains ($L \sim N$), moderately fast for random networks of the Erd\"os-R\'enyi type ($L \sim \ln N$), the small-world networks ($L \lesssim \ln N$) and for the Bar\'abasi-Albert networks ($L \sim \ln N / \ln \ln N$)~\cite{albert02}, slowly for the ultrasmall networks ($L \sim \ln \ln N$)~\cite{cohen03}, while for densely connected networks it can roughly be independent of $N$. For our networks, values of this quantity (for the complete texts) belong to the interval: $2.7 \le L \le 3.8$, which generally falls into the small-world networks regime. However, the asymptotic behavior of $L(N)$ is significantly different from the small-world one since for $N \gg 1$, $L$ is decreasing function of $N$ (Figure~\ref{fig::path.nodes}). This is because, for large $N$, the vocabulary volume $V$ (here, $V = N$) used for writing a text grows much more slowly than the length $T$ of the text (which is expressed by a general relation $V \sim T^{\delta}$, where $0.4 \lesssim \delta \lesssim 0.6$ - the so-called Heaps law) and this leads to increasing density of edges.

The clustering coefficient $C$ for an undirected binary network is expressed by:
\begin{equation}
C = {1 \over N} \sum_{i} {\sum_{j,m} a_{ij} a_{jm} a_{mi} \over k(k-1)},
\end{equation}
where $a_{pq}$ are binary edges (equal to 1 for the existing edge and 0 otherwise). Its value is typically small for the Erd\"os-R\'enyi networks ($C \sim N^{-1}$) and for the Bar\'abasi-Albert networks ($C \sim N^{-0.75}$)~\cite{albert02}, while it is large (and independent of $N$) for the small-world networks of the Watts-Strogatz type~\cite{watts98}.

\begin{figure}
\centerline{\psfig{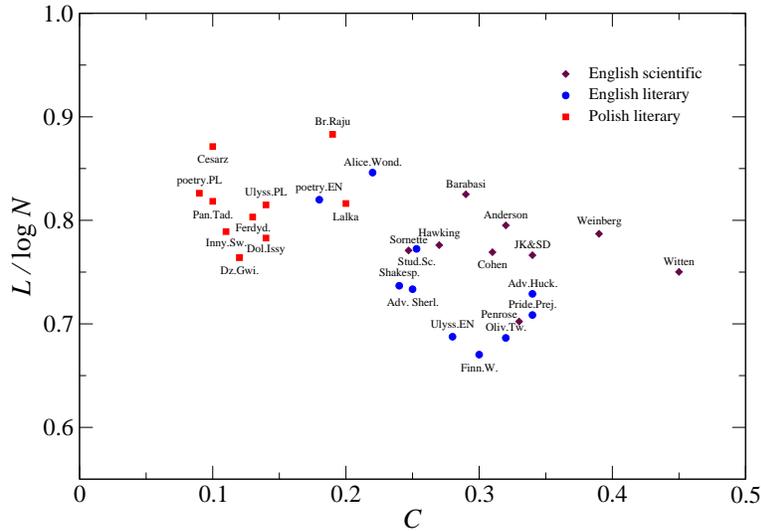}}
\caption{The clustering coefficient $C$ vs. the average shortest path length $L / \log N$ for the texts considered in this work. Different groups of texts are denoted by different symbols.}
\label{fig::path.vs.coeff}
\end{figure}

For all the texts except the highly mathematical papers by E.~Witten and S.~Weinberg, values of $C$ are in the interval $0.09 \le C \le 0.34$, but unlike the shortest path length, here we observe rather clear separation between Polish and English novels (including the Shakespeare's drama): $0.10 \le C \le 0.20$ for the Polish ones and $0.22 \le C \le 0.34$ for the English ones. The poetry in both languages tends to have smaller clustering coefficient than any other considered piece of text: $C_{\rm PL}=0.09$ and $C_{\rm EN}=0.18$. These results can be seen in a scatter plot in Figure~\ref{fig::path.vs.coeff}, which shows values of the clustering coefficient and the shortest path length for all the texts considered in this work. Indeed, the Polish and the English texts occupy different regions of the ($C$,$L$) plane with the Polish ones being characterized by smaller $C$. For English, one can also see that scientific texts may have different properties than the literary texts, especially if they contain much mathematics (Witten, Weinberg). If they are more descriptional than mathematical, their properties can resemble the properties of literary texts (Sornette, Hawking, Penrose).

\begin{figure}
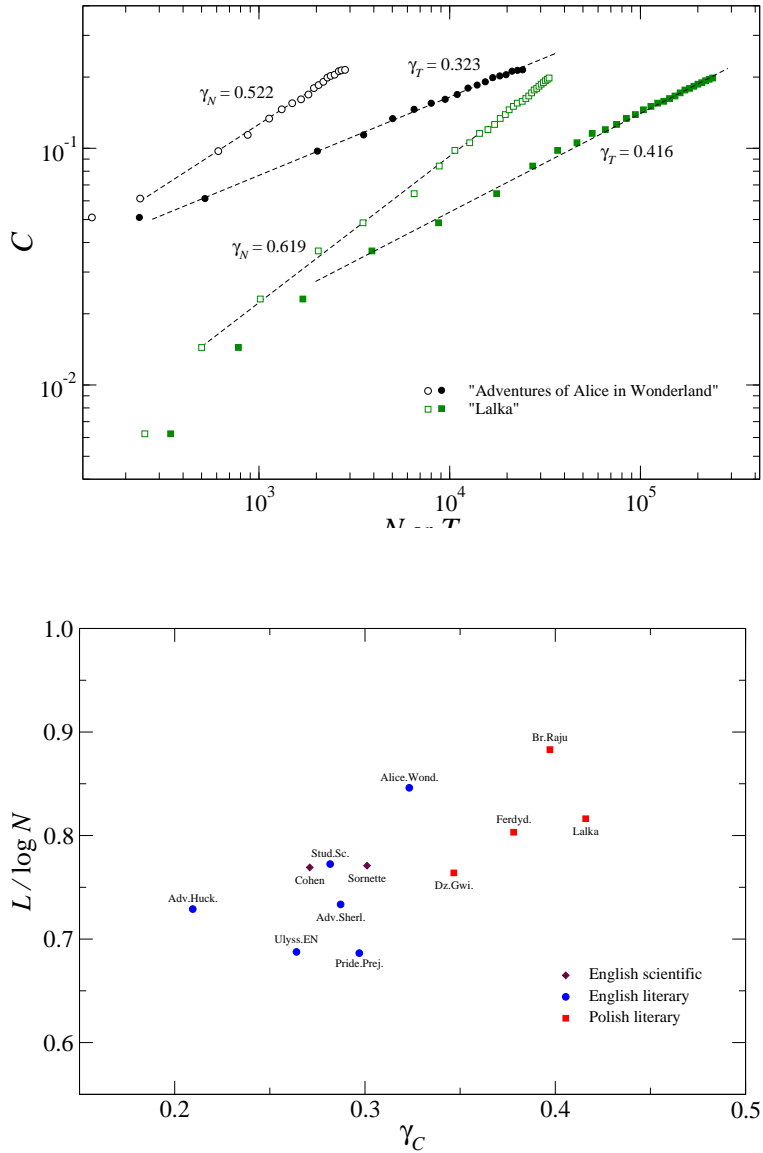

\centerline{\psfig{file=fig9a.eps,width=10cm}}

\vspace{1.0cm}
\centerline{\psfig{file=fig9b.eps,width=10cm}}
\caption{(a) The clustering coefficient $C$ dependence on the number of nodes $N$ and the text length $T$ for two exemplary texts with the observed scaling behavior $C \sim N^{\gamma_N}$ or $C \sim T^{\gamma_T}$. The scaling indices $\gamma_N$ and $\gamma_T$ are shown together with the corresponding least-squares fits. (b) The scaling index $\gamma_T$ vs. the average shortest path length $L$ for those texts for which $C(T)$ was at least partially power-law. Different groups of texts are denoted by different symbols.}
\label{fig::clust.power.law}
\end{figure}

It is interesting to note that unlike other network types mentioned above, for our adjacency networks, $C$ is an increasing function of $N$ and, typically, its dependence is either power-law $C \sim N^{\gamma_N}$ (at least for large values of $N$) with the scaling index $\gamma_N < 1$ or not far from power-law. A similar power-law behavior can be seen for $C(T)$, where $T$ is the text length, but this is not surprising due to the already-mentioned Heaps law: $N \sim T^{\delta}$. Figure~\ref{fig::clust.power.law}a shows two examples of such power-law behavior for one English and one Polish text. As it was the case with $C$, the English and the Polish texts have distinct values of $\gamma_T$, with the latter being significantly larger. Calculated values of this index for the texts that show clear power-law dependence of $C(T)$ are collected in Figure~\ref{fig::clust.power.law}b. Although $\gamma_T$ cannot be estimated for all the texts from Figure~\ref{fig::path.vs.coeff}, the separation of the languages is clear also here. 

The observed behavior of $L(N)$ and $C(N)$ suggests that the networks considered in this work have their own specific structure that clearly distinguishes them from the most well-known network structures. They cannot be counted as small-world networks even though the average characteristic path length is relatively short. They also differ from the Bar\'abasi-Albert networks despite the fact that some of they show the scale-free structure and from the random (Erd\"os-R\'enyi) ones.

\section{Conclusions}

We presented several results from our quantitative study of statistical and network properties of literary and scientific texts written in two substantially different languages: English and Polish. We transformed the text samples into word-adjacency networks defined by the nodes representing individual words and the edges representing pairs of directly neighbouring words. For the majority of the studied literary texts in both languages, the corresponding networks revealed the scale-free structure, while this was rarely the case for the scientific texts. We also showed that there are differences in node degree distributions between prose and poetry, especially in English. Poetry has a detectable steeper distribution's slope than has prose. The slope for scientific texts is even steeper than for poetry, but this can be explained by typically poorer vocabulary in the former case. Despite these differences, all the network representations of texts were hierarchical with a few important hubs and the majority of less important nodes. No qualitative and quantitative difference between the languages was noticed in this respect. This picture changed completely if we looked at other network statistics like the clustering coefficient and the average shortest path length. The English texts appear to possess more clustered structure, while the Polish ones were less clustered. This result was attributed to differences in grammar of both languages, which was also indicated in the Zipf plots. Our results suggest that the word-adjacency networks cannot fully be described by any of the Erd\"os-R\'enyi, the Watts-Strogatz, and the Bar\'abasi-Albert models even though these networks exhibit certain characteristics of the latter two models. Such networks will be a subject of our forthcoming study.

\end{document}